\documentclass[aps,prb,twocolumn,amsmath,amssymb,nofootinbib,superscriptaddress]{revtex4-2}

\usepackage{amsmath}
\usepackage{amssymb}
\usepackage{graphicx}
\usepackage{array}
\usepackage{bm}
\usepackage{braket}
\usepackage{booktabs}
\usepackage{dsfont}
\usepackage{dcolumn}
\usepackage{esint}
\usepackage{epstopdf}
\usepackage{mathrsfs}
\usepackage{tabu}
\usepackage{enumitem}

\newcommand{\bt}{\textbf}

\newcommand{\appropto}{\mathrel{\vcenter{
  \offinterlineskip\halign{\hfil$##$\cr
    \propto\cr\noalign{\kern2pt}\sim\cr\noalign{\kern-2pt}}}}}

\usepackage[colorlinks=true,citecolor=green,linkcolor=blue]{hyperref}
\usepackage{color}

\begin{document}

\title{Experimental review on Majorana zero-modes in hybrid nanowires}

\author{Ji-Bang Fu}
\affiliation{Institute for Quantum Information \& State Key Laboratory of High Performance Computing, College of Computer Science and Technology, NUDT, Changsha 410073, China}
\author{Bin Li}
\affiliation{Institute for Quantum Information \& State Key Laboratory of High Performance Computing, College of Computer Science and Technology, NUDT, Changsha 410073, China}
\author{Xin-Fang Zhang}
\affiliation{Institute for Quantum Information \& State Key Laboratory of High Performance Computing, College of Computer Science and Technology, NUDT, Changsha 410073, China}
\author{Guang-Zheng Yu}
\affiliation{Institute for Quantum Information \& State Key Laboratory of High Performance Computing, College of Computer Science and Technology, NUDT, Changsha 410073, China}
\author{Guang-Yao Huang}
\affiliation{Institute for Quantum Information \& State Key Laboratory of High Performance Computing, College of Computer Science and Technology, NUDT, Changsha 410073, China}
\author{Ming-Tang Deng}
\email[Corresponding author:]{mtdeng@nudt.edu.cn}

\begin{abstract}
As the condensed matter analog of Majorana fermion, the Majorana zero-mode is well known as a building block of fault-tolerant topological quantum computing. This review focuses on the recent progress of Majorana experiments, especially experiments about semiconductor-superconductor hybrid devices. We first sketch Majorana zero-mode formation from a bottom-up view, which is more suitable for beginners and experimentalists. Then, we survey the status of zero-energy state signatures reported recently, from zero-energy conductance peaks, the oscillations, the quantization, and the interactions with extra degrees of freedom. We also give prospects of future experiments for advancing one-dimensional semiconductor nanowire-superconductor hybrid materials and devices.
\end{abstract}

\date{\today}

\maketitle

\section{Introduction}

\begin{figure}[t]
	\centering
	\includegraphics[width=8 cm]{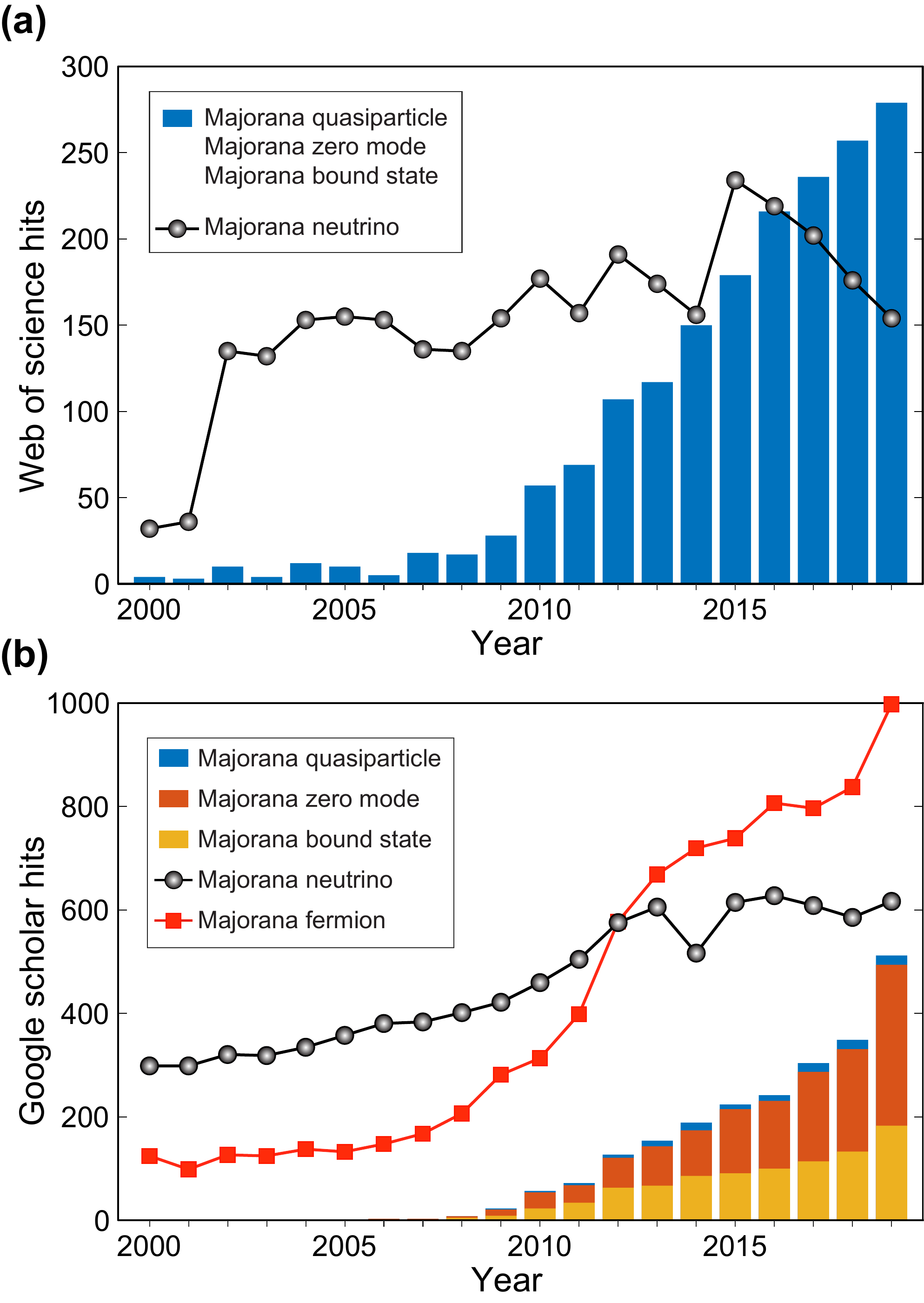}
	\caption{\label{Fig1} Majorana fermions related research trend over the past 20 years. (a) Topic hits from the \emph{Web of Science} with different keywords. (b) Keyword hits from the \emph{Google Scholar}.}
	\label{fig:fig1}
\end{figure}

Majorana zero-modes have been intensively studied in recent years (see the research trend shown in Fig.~\ref{Fig1}) owing to their potential application in fault-tolerant quantum computing. There have already been many review papers about  Majoranas~\cite{Wilczek2009, Leijnse2012semi, Elliott2015, Sarma2015, Aguado2017, Lutchyn2018, Prada2020, Schuray2020}, mainly on Majorana theories. This review will briefly introduce recent experimental developments of one-dimensional Majorana devices, aiming to give a comprehensive overview and understanding of one-dimensional Majorana experiments. 

Despite a series of observed signatures among a wide range of platforms, there clearly exist disputes over the nontriviality of the observations. To date, a common consensus of this field is that the concrete experimental evidence of the non-Abelian Majorana zero-mode is still missing. This experimental review will focus on the reported signatures of Majorana zero-modes and analyze why they are still not solid enough to demonstrate non-Abelian operations.

First of all, we will briefly introduce some fundamental conceptions of the Majorana theories, and here we start from anyons. Anyons, particles defined in two-dimensional systems with fractional statistics~\cite{Wilczek1982, Wilczek1990}, have been predicted to be able to perform quantum computation since the early 1990s~\cite{Castagnoli1993}. It was further pointed out that anyon-based quantum computing is naturally fault-tolerant (without a large overhead of quantum error-correction) if the anyons' braiding group is non-Abelian~\cite{Kitaev2003}. This hardware-level fault tolerance is endowed by the so-called topological protection~\cite{Nayak2008, Pachos2012}. 

The quantum information encoded in non-Abelian anyons is only accessible when the anyons are brought together. Unitary operations, in the form of intertwining anyons that are wide apart, remain robust to disturbances, provided that the perturbations are weaker than the energy gap separating the excitations from the degenerate ground state.

Majorana zero-mode (or Majorana bound state, Majorana quasiparticle) is one of the simplest realizations of the non-Abelian anyons (Ising type, specifically). It is a condensed matter analog of Majorana particle, an elementary fermion with its antiparticle being itself~\cite{Majorana1937, Cardani2019}. It should be noted that the terminology ``Majorana" only means this quasiparticle has a real fermion operator. Majorana first came into condensed matter physics as an operator elucidating the Moore-Read Pfaffian states in fractional quantum Hall systems~\cite{Moore1991}, and it was then used to refer to the zero-energy modes in half-quantum vortices of chiral superconductors~\cite{Read2000, Senthil2000} or superfluids~\cite{Volovik1999, Ivanov2001}. The chiral superconductivity/superfluidity possesses an order parameter of $p_x$+$ip_y$ symmetry, and hence, they are also called the $p$-wave superconductivity/superfluidity. Although there are a few candidates of intrinsic $p$-wave superfluids (like \textsuperscript{3}He-A superfluid phase~\cite{Volovik2003}) and superconductors (e.g., Sr\textsubscript{2}RuO\textsubscript{4}, Cu\textsubscript{x}Bi\textsubscript{2}Se\textsubscript{3}, CePt\textsubscript{3}Si, see the review paper Ref.~\cite{Sato2017}), strong evidence of $p_x$+$ip_y$-pairing is still missing in these systems. 

In the meantime, the introduction of topological insulators has given birth to artificial $p$-wave superconductors. In these engineered two-dimensional $p$-wave superconductors, the spin-singlet pairing potential can be induced from an $s$-wave superconductor to a spin-polarized (i.e., spinless) hosting material. Typically, spin-polarized systems (such as ferromagnet) are hostile environments for superconductivity. However, if the hosting material's band structure is topological-insulator-like, where the material possesses a single Dirac cone with a helical spin texture, a finite spin-triplet order parameter can be established at the hybridization interface. With a strong spin-orbit locking effect and a unique band structure, a topological insulator in proximity to an $s$-wave superconductor is thus a viable Majorana zero-mode platform~\cite{Fu2008}. It was then proven that a semiconductor with strong spin-orbit interaction could play a similar role to the topological insulator in the presence of a magnetic insulator~\cite{Sau2010} or an in-plane Zeeman field~\cite{Alicea2010}.

Although Majorana zero-mode was originally envisioned in two-dimensional $p$-wave superconductors, it can also be supported at the phase boundaries (endpoints or domain walls) of one-dimensional systems [Fig.~\ref{Fig2}(a)]. In fact, the one-dimensional $p$-wave superconductor is one of the earliest and theoretically the most straightforward Majorana-harboring proposals. The corresponding toy model is presented as the so-called Kitaev chain model~\cite{Kitaev2001}. One of the implementations geometrically similar to the Kitaev model is a one-dimensional semiconductor-superconductor nanowire that can support Majoranas at its ends~\cite{Lutchyn2010, Oreg2010}. Without the hazard of fusion, the adiabatic braiding operations of Majorana zero-mode in a nanowire can be done via a bypassing channel in the wire middle, i.e., a ``T"-junction geometry~\cite{Alicea2011}. Henceforth, the topological quantum computation can be realized in nanowire networks assembled by such ``T"-junctions~\cite{Karzig2017, Schrade2018}.

Experimentally pursuing Majorana zero-mode almost immediately began after the theoretical ideas were proposed, and a large number of observations of possible Majorana signatures have been reported. The two most extensively utilized experimental techniques are the transport measurements in semiconductor-superconductor hybrids~\cite{Mourik2012, Deng2012, Rokhinson2012, Das2012, Churchill2013, Finck2013, Deng2014, Albrecht2016, Deng2016, Suominen2017, Nichele2017, GulO2018, Vaitiekenas2018, Deng2018, Shen2018, Fornieri2019, Ren2019, Grivnin2019, Anselmetti2019, Desjardins2019, Vaitiekenas2020, Vaitiekenas2020-2, ShenJ2020, PanD2020} and the scanning-tunneling-microscope based measurements on various materials~\cite{Perge2014, Xu2015, Sun2016, Pawlak2016, Feldman2017, Jeon2017, Zhang2018s, Wang2018, Liu2018, Kong2019, Yuan2019, Menard2019, Chen2019cpl, Zhu2020, Wang2020, Chen2020, Manna2020, Liu2020, Chen2020nanoL}. As we mentioned above, this review will focus on the first branch, though some relevant experiments of the second branch will also be discussed. One can find more details of the second branch in the review papers Refs.~\cite{Kong2019NSR, Li2019, Zheng2019}. Other Majorana quasiparticle experiments, such as 4$\pi$-periodic supercurrent~\cite{Bocquillon2017, Deacon2017, Laroche2019}, chiral Majorana edge states~\cite{He2017}, and the debates about those observations~\cite{Dartiailh2021, Kayyalha2020},will not be addressed in detail here.

\section{Majorana zero-mode: a bottom-up view}
\begin{figure*}[t]
\centering \includegraphics[width=18 cm]{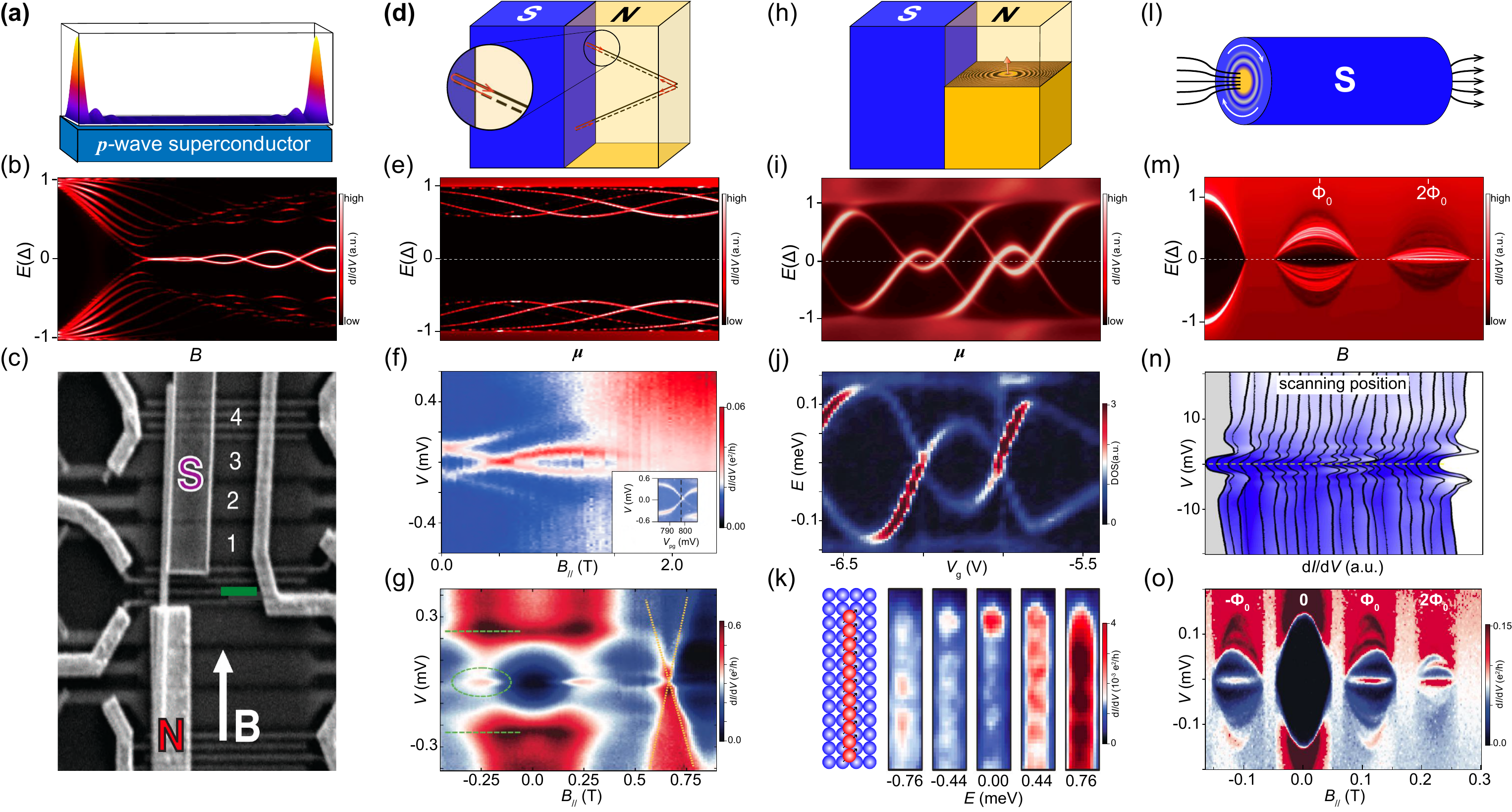}
\caption{\label{Fig2} Sub-gap bound states in superconducting systems.
(a) Schematic of Majorana wave function in a spinless one-dimensional \emph{p}-wave superconductor. The wave function demonstrates an end-to-end nonlocal distribution.
(b) Calculated spectrum of sub-gap state in a Majorana wire, as a function of the magnetic field. Majorana zero-mode emerges as the superconducting gap closes and reopens at the phase transition point. Data reproduced from the supplementary information of Ref.~\cite{Deng2016}. 
(c) Scanning-electron-microscope image of a superconductor-semiconductor nanowire device. Adopted from Ref.~\cite{Mourik2012}. 
(d) Schematic of the formation of Andreev bound state.
(e) Calculated spectrum of trivial Andreev bound states in a one-dimensional superconductor-semiconductor system. Data reproduced from the supplementary information of Ref.~\cite{Deng2016}.
(f) Measured trivial Andreev bound state spectrum evolving in the magnetic field. Adopted from Ref.~\cite{Lee2014}. 
(g) Measured zero-energy conductance peak as a signature of Majorana bound state. Adopted from Ref.~\cite{Mourik2012}. 
(h) Schematic of Yu-Shiba-Rusinov sub-gap states in a superconducting-magnetic impurity system.
(i) Calculated trivial Yu-Shiba-Rusinov state in a spinful Josephson quantum dot setup.
(j) Measured trivial Yu-Shiba-Rusinov spectrum in a carbon nanotube quantum dot sandwiched by two superconductor leads. Data in (i) and (j) are reproduced from Ref.~\cite{Pillet2010}. 
(k) Scanning-tunneling-microscope spectrum of a ferromagnetic Yu-Shiba-Rusinov chain on a superconductor, where a zero-energy located at the chain end is visible. Adopted from Ref.~\cite{Perge2014}. 
(l) Schematic of the vortex Caroli-de Gennes-Matricon state. 
(m) Calculated spectrum of trivial Caroli-de Gennes-Matricon states for a semiconductor-superconductor core-shell model. Data reproduced from Ref.~\cite{Vaitiekenas2020}. 
(n) Measured trivial Caroli-de Gennes-Matricon vortex states in a high-$T_c$ superconductor. Adopted from Ref.~\cite{Chen2018}. 
(o) Measured vortex bound state spectrum of a semiconductor-superconductor core-shell nanowire, where a zero-energy state appears with an odd winding number. Data reproduced from Ref.~\cite{Vaitiekenas2020}. 
Some adopted figures are unified to the same colormaps for easier comparison.} 
\end{figure*}
The formation of Majorana zero-mode is a conclusion of phase transition with the change in band topology. We can use a less rigid  ``bottom-up" picture for the semiconductor-superconductor hybrid system to help understand how the Majorana zero-modes are established.

In the above section, we have learned that a semiconductor-superconductor hybrid nanowire is a possible Majorana platform. In order to create Majorana zero-modes in such a system, semiconductor nanowires with large spin-orbit interactions need to be electronically contacted to superconductors [see the scanning-electron-microscope image of a typical device in Fig.~\ref{Fig2}(c)]. Besides the spin-orbit field, an external magnetic field (from a ferromagnet or a coil) is also required to break the time-reversal symmetry. If the external magnetic field is (at least partly) orthogonal to the spin-orbit field, a helical gap can be opened around $\mu=0$, with $\mu$ the semiconductor chemical potential.  Practically, $\mu$ can be tuned by electric gates near the nanowire. For a single-channel nanowire, if the Zeeman energy $V_z$ fulfills $V_z \geq \sqrt{\mu^2+\Delta^{\prime2}}$, Majorana zero-modes can emerge at the end of the wire, with $\Delta^{\prime}$ the induced gap [Fig.~\ref{Fig2}(b)]. A tunneling probe (a normal metal in most cases) with a moderate opaqueness is contacted to the system to detect the formed Majorana zero-modes.

Amidst the Majorana ingredients, the superconducting proximity effect plays a critical role in the engineered
topological superconductors. It allows the BCS wavefunction to ``leak" out the superconductor into a non-superconducting material in its close vicinity, with some inherited properties of the hosting materials. Since the proximity effect in low-dimension systems always appears along with the formation of Andreev bound states, which have a deep connection to Majorana bound states, we will first elaborate on Andreev bound states in-depth below.

Electrons in low-dimensional systems are strongly confined, and the confinements give rise to discrete levels with quantized energies. To distinguish these confined states from their superconducting counterparts, we will refer to them as normal bound states in the rest of this review. Just as the word ``bound" indicates, the particles occupying these bound states are coherently bounced back and forth by the system boundaries. In a superconductor-normal conductor hybrid system, the so-called Andreev reflection, instead of the normal reflection, occurs when a low-energy incident particle (or hole) travels from the normal conductor into the superconductor. Because of the lack of available states in the superconductor gap $E<|\pm \Delta|$, the incident particle (hole) cannot directly propagate into the superconductor. Consequently, the superconductor will retro-reflect (Andreev reflection) a time-reversed hole (particle) with opposite momentum and spin into the normal conductor. In this way, the superconductor gains (loses) a Cooper pair from (to) the normal conductor. Electron states confined by Andreev reflection are hence called Andreev bound states [see Fig.~\ref{Fig2}(d)], which are essentially the extension of the superconducting pairing potential $\Delta (r)$ out of the superconductor.

Since Andreev reflection is a phase-coherent two-electron process, the energy of Andreev bound states $\zeta$ can be modified by adjusting the boundary phase difference (applying a phase-bias across the superconductors) or the picked-phase on a round trip [varying the incident particle momentum via the chemical potential control, see Fig.~\ref{Fig2}(e)]~\cite{Houten1991, Bena2011}. For example, the phase dependence of Andreev bound state energy $\zeta$ in an SNS structure gives $\zeta  \propto  \sqrt{1-\tau ^2 \sin (\phi /2) }$, with $\tau$ the transmission coefficient and $\phi$ the phase difference between the two superconductors~\cite{Beenakker1991}. Here,  $\zeta$ always has a non-zero value except for unit transmission. However, the Andreev bound states can split in the presence of the Zeeman field and undergo a spin-singlet$\rightarrow$spin-doublet phase transition. The lower-energy branch will cross zero-energy at the phase transition point [see Fig.~\ref{Fig2}(f)]. 

Some trivial sub-gap states can stay at zero-energy with zero Zeeman field. For instance, Yu-Shiba-Rusinov states, the bound states formed in superconductors with local magnetic moments~\cite{Yu1965, Shiba1968, Rusinov1969}, can have zero-energy when the magnet-screening effect is tuned to a sweet spot [see Figs.~\ref{Fig2}(h-i)].

\begin{figure}[t]
\centering \includegraphics[width=8.5 cm]{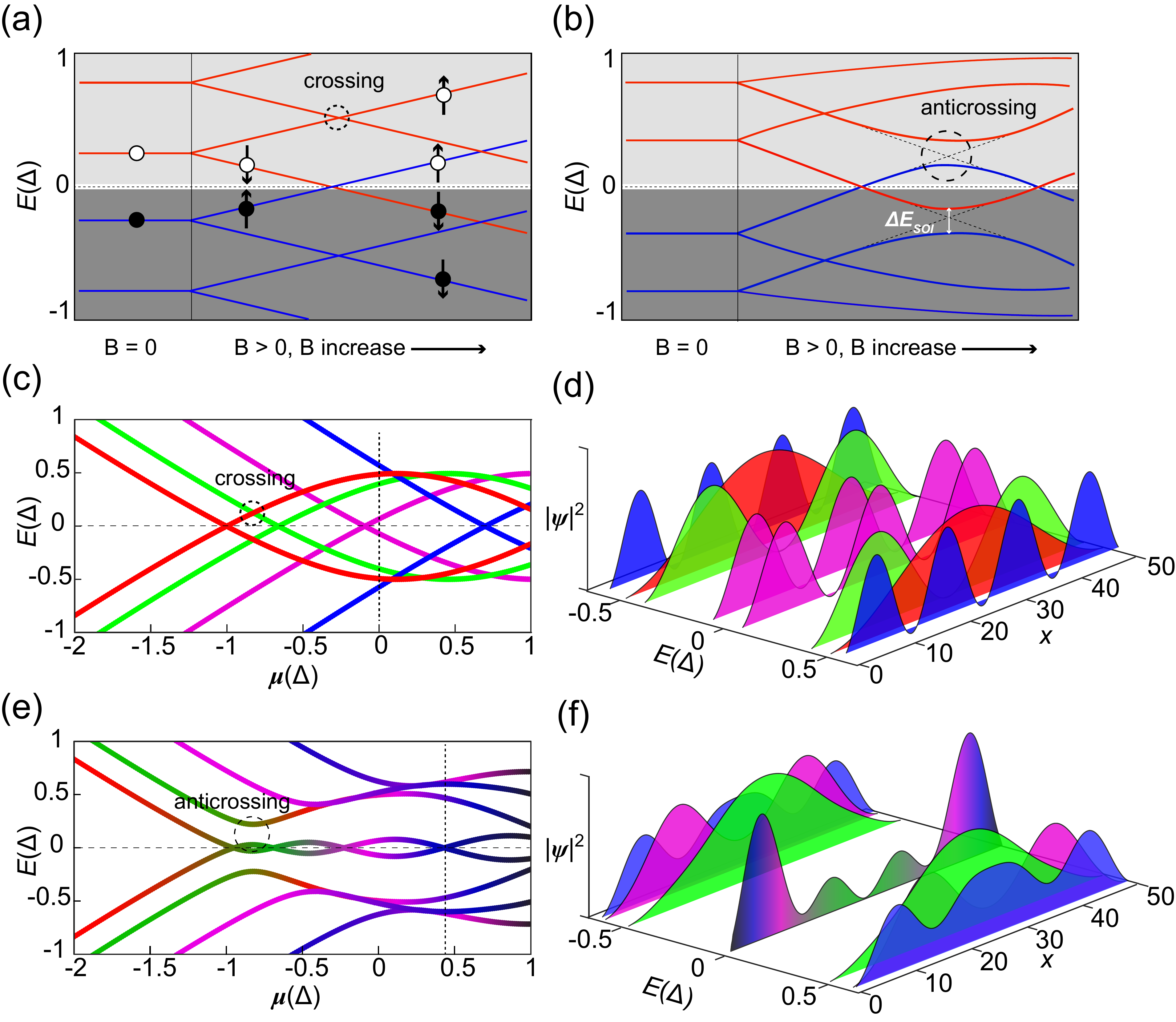}
\caption{\label{Fig3} Coalescing Andreev bound states into Majorana bound state. (a) Schematic view of the Andreev bound state evolution in the magnetic field. (b) Same with (a), but with a spin-orbit interaction $\Delta E_{\rm SOI}$. (c) Calculated spectrum of Andreev bound states without spin-orbit interaction in a one-dimensional hybrid wire as a function of chemical potential. (d) Wave functions of the Andreev bound states shown in (c). Without the mixing effect from spin-orbit interaction, the Andreev bound states evolve independently. (e-f) The same with (c-d), but with a finite spin-orbit interaction. The presence of spin-orbit interaction mixes trivial Andreev bound states with different wave vectors and gives rise to a new state whose wave function locates at the end of the one-dimensional hybrid wire.} 
\end{figure}

The Yu-Shiba-Rusinov states can continuously mutate to ordinary Andreev bound states by varying the local spin arrangement~\cite{Meng2015}. Therefore,  the Yu-Shiba-Rusinov states are also dubbed a kind of Andreev bound states in the context of low-energy excitation in superconductors. Generally speaking, the low-energy excitation in an Abrikosov vortex, i.e., Caroli-de Gennes-Matricon state~\cite{Caroli1964}, is another special kind of Andreev bound state [see Figs.~\ref{Fig2}(l)-(n)]. 

Andreev bound states (including Yu-Shiba-Rusinov states and Caroli-de Gennes-Matricon states) with a broken time-reversal symmetry can be coalesced into Majorana bound states in the presence of a moderate spin-orbit field. Here, we use the word ``coalesce" to emphasize that Majorana bound states can directly evolve from Andreev bound states.

Corresponding to discrete normal bound states, discrete Andreev bound states can form in a one-dimensional semiconductor-superconductor hybrid wire with a finite length. Those states have different wave vectors and are spin-degenerate at zero magnetic field. As an external magnetic field is applied, each of the Andreev bound states spits into two branches with opposite spins. In the scenario without any interaction between the individual Andreev bound states, one of the two spin-branches (i.e., the inner branch) evolves towards zero-energy as the magnetic field increases. If the bulk superconducting gap can survive at an even higher magnetic field, the inner Andreev bound state will cross zero energy.  In some low-dimensional systems with few states, crossing zero-energy indicates a phase transition from a spin-singlet ground state to a spin-doublet ground state. The crossed Andreev bound state will further evolve monotonously and crosses with higher energy levels [see Fig.~\ref{Fig3}(a)]. 

However, if the system has a large spin-orbit interaction, level anticrossing between different levels occurs at finite energy [Fig.~\ref{Fig3}(b)]. It seems that the split low-energy Andreev bound states are pushed back to zero-energy~\cite{Bommer2019}. One can imagine that the low-energy Andreev bound states can be pinned at zero-energy if there are enough repulsive high-energy states. Moreover, the spin-orbit interaction mixes the wave functions of two states at the level anticrossing point. The state pinned at zero energy effectively comprises the wave vector components from all of the states that it couples to. As shown in Fig.~\ref{Fig3}(f), from the Fourier transformation point of view, the wave function of the zero-energy state can support $\delta$-function like singularity points at system boundaries. If this picture holds, we get Majorana bound state at those boundaries. This bottom-up view reveals how the Andreev bound states coalesce into Majorana bound states in the presence of the Zeeman field and the spin-orbit field.

From the coalescing Andreev bound state point of view, a long one-dimensional system with strong spin-orbit interaction can provide enough interacting states to pin the lowest-energy state at zero-energy and guarantee a more $\delta$-function  like wave function of the Majorana state, i.e., a pair of well-decoupled Majorana states. We will come back to this picture in the following sections when discussing different aspects of Majorana signatures.

\section{Zero-energy conductance peak}
\begin{figure*}
\centering \includegraphics[width=18 cm]{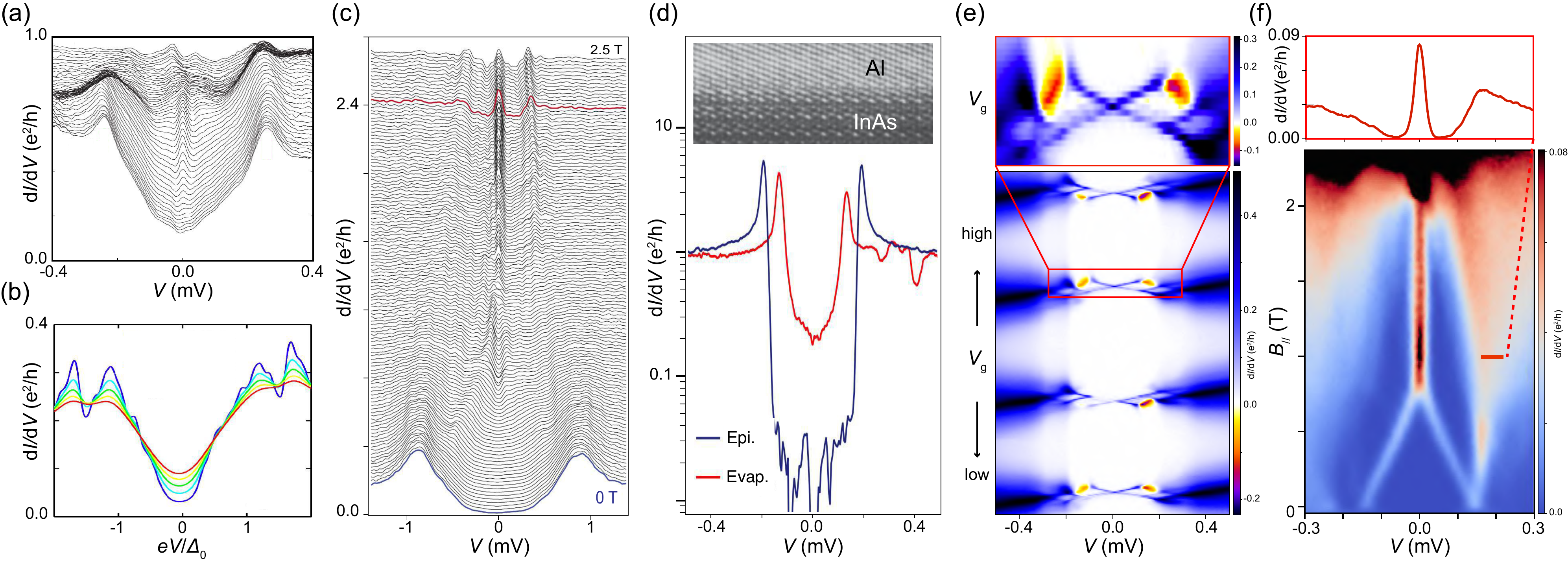}
\caption{\label{Fig4} Zero-energy conductance peaks and soft-gap problem.
(a) Tunneling conductance spectrum of an InSb nanowire-NbTiN device. Sizable conductance remains in the superconducting gap, even at zero magnetic field. Data reproduced from Ref.~\cite{Mourik2012}. 
(b) Calculated superconducting gap softness induced by semiconductor-superconductor interface inhomogeneity. Adopted from Ref.~\cite{Takei2013}. 
(c) Tunneling conductance spectrum measured for an InSb nanowire-NbTiN device with an improved interface quality. Adopted from Ref.~\cite{GulO2018}. 
(d) Gap hardness comparison between InAs/Al (evaporated) hybrid wire and InAs/Al (epitaxial grown) hybrid wire. Data is adopted from Ref.~\cite{Chang2015}. Inset: Transmission-electron-microscope image of the full-epitaxial InAs/Al interface, adopted from Ref.~\cite{Krogstrup2015}. 
(e) Observed negative differential conductance from a full-epitaxial InAs/Al Coulomb island, which can be attributed to the low quasiparticle relaxation frequency. Data reproduced from Ref.~\cite{Higginbotham2015}, with a modified colormap to highlight the negative differential conductance. 
(f) Zero-energy conductance peak emerged in a hard superconducting gap. Adopted from Ref.~\cite{Deng2016}.} 
\end{figure*}

According to theory predictions, Majorana bound state will emerge in the mid-gap of one-dimensional $p$-wave superconducting system and give rise to a zero-energy conductance peak if measured by a nearby normal tunneling probe. The first zero-energy conductance peak signature was reported in Ref.~\cite{Mourik2012}, as shown in Fig.~\ref{Fig4}(a) [from the device shown in Fig.~\ref{Fig2}(c)]. Similar signatures were also observed later in Refs.~\cite{Deng2012, Das2012, Churchill2013, Finck2013, Deng2014}. However, in all of these experiments, the zero-energy conductance peaks are accompanied by non-zero sub-gap conductance backgrounds induced by sizable disorders, i.e., the so-called soft-gap problem. Soft-gap could be fatal to Majorana's application. This is because Majorana-based quantum information is normally coded in the charge/parity basis, i.e., the parity of the fermionic quasiparticle number of a Majorana qubit. The soft-gap will behave like a quasiparticle bath, which can flip the parity of Majorana zero modes and violate the topological protection.

Several effects can cause the soft-gap problem, such as impurities (magnetic/nonmagnetic), BCS broadening effects (temperature), and the nonuniformity of the interface between the semiconductor and the superconductor. Reference~\cite{Takei2013} theoretically investigated soft gaps induced by various effects [Fig.~\ref{Fig4}(b)] and concluded that interface inhomogeneities could be the main reason for the soft gaps observed in Refs.~\cite{Mourik2012, Deng2012, Das2012, Churchill2013, Finck2013, Deng2014}. Therefore, improving the semiconductor-superconductor interface quality is important for observing hard-gap Majorana.

The main obstacle to forming a uniform interface is the oxidization of the semiconductor surface. Usually,  the semiconductor and the superconductor are fabricated separately. The semiconductor surface will be oxidized in the ambient environment and form an insulating layer. Thus, the surface cleaning process is needed to remove the oxidized layer before evaporating the superconductor on it via either dry etching (plasma ion etching) or wet etching. Inevitable implantation of defects, disorders, or dopant atoms will be induced by ion bombardment or chemical reaction. Extensive efforts for minimizing these damages have been made. For example, optimizing the InSb nanowire surface retreatment recipe can result in a larger and much cleaner induced gap in an InSb-NbTiN hybrid system~\cite{GulO2018}. 

A further breakthrough comes from material synthesizing. Reference~\cite{Krogstrup2015} reported an \emph{in-situ} semiconductor-superconductor growth method, in which superconductor (Al) were directly grown on semiconductor nanowire (InAs) without breaking the vacuum. Therefore, surface oxidization/etching and the implanted interface damage were avoided. The atom-by-atom lattice matching guarantees a highly uniform interface. As shown by the tunneling spectra in Fig.~\ref{Fig4}(d), the induced superconductor gap in the full epitaxy hybrid nanowire device (the blue curve) is much harder than the gap of the evaporated superconductor nanowire device (the red curve). With reduced residual sub-gap states, quasiparticle relaxation frequency can be greatly suppressed. In Ref.~\cite{Higginbotham2015}, the sub-gap negative differential conductance caused by the quasiparticle blocking effect is used to measure the quasiparticle relaxation/poison frequency [Fig.~\ref{Fig4}(e)]. This frequency can be as low as $\sim$10\textsuperscript{2} Hz, thanks to the hard gap of the full-epitaxy hybrid nanowire. Note that the quasiparticle poison frequency also strongly depends on the tunneling rate to the two normal metal contacts~\cite{Albrecht2017}. The hard gap can also persist in a strong parallel magnetic field~\cite{Deng2016}, and robust zero-energy modes emerge from the clean background [Fig.~\ref{Fig4}(f)].

However, as noticed in Ref.~\cite{Vaitiekenas2018}, the induced gap of the full-epitaxy hybrid nanowire in the high electron-density regime (where the effective $g$-factor is large) is still not hard enough. Besides the soft-gap problem, material metallization~\cite{Antipov2018, Mikkelsen2018, Reeg2018}, fabrication induced quantum dot or non-uniform potential profile~\cite{Liu2017, Chen2019, ReegPRB2018}, and magnetic field induced quasiparticle poisoning~\cite{Higginbotham2015, Albrecht2017} also severely hinders the formation of highly decoupled Majorana bound states and their application. Thus, more innovations from material science and nanofabrication technology are required. 

\section{Majorana oscillations}

\begin{figure*}
\centering \includegraphics[width=18 cm]{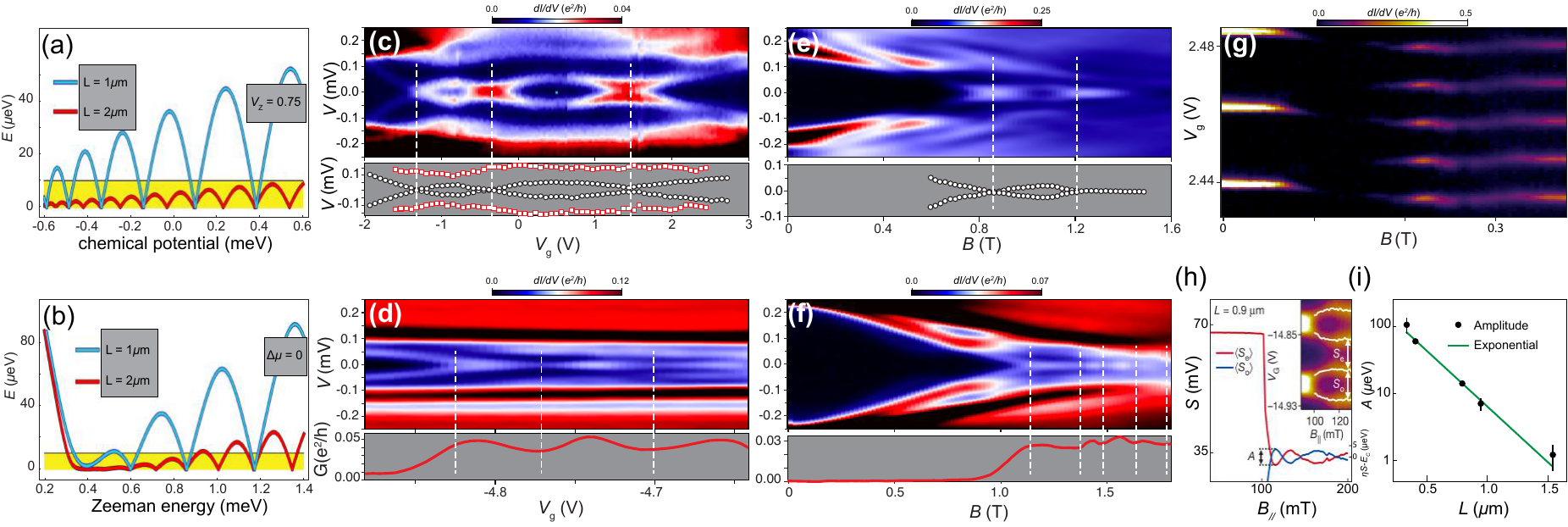}
\caption{\label{Fig5} Signature of Majorana oscillations.
(a) and (b) Calculated oscillations of Majorana modes with chemical potential and Zeeman field, respectively. Adopted from Ref.~\cite{Sarma2012PRB}. 
(c) and (d) Measured sub-gap bound states in epitaxial InAs/Al nanowires, with oscillating energy and conductance as a function of gate voltage.
(e-f) Same as in (c-d), but oscillating with the magnetic field. Data in (c) is reproduced from Ref.~\cite{Deng2016}, and data in (d-f) are reproduced from Ref.~\cite{Deng2018}. 
(g)-(i) Measured Majorana-island geometry data. The energy of the lowest sub-gap state can be extracted via peak-to-peak distance. The energy oscillation of the sub-gap bound states shows an exponential decay as the island length increases. Adopted from Ref.~\cite{Albrecht2016}.} 
\end{figure*}

As more and more emergent zero-energy conductance peaks in superconductor-semiconductor hybrid systems (and in other platforms) are observed, it is realized that some non-Majorana physics can also give rise to similar conductance peaks. These Majorana-mimicking mechanisms include Kondo effects~\cite{Lee2012}, weak-antilocalization~\cite{Pikulin2012}, trivial Andreev/Yu-Shiba-Rusinov/Caroli-de Gennes-Matricon bound states~\cite{Lee2014, Chen2019, Scherubl2020, Kong2019}. Notably, the difference between trivial sub-gap states and partially coupled Majorana bound states is subtle. Actually, as we will discuss later, it expands a continuum between completely topologically trivial (no topological protection) bound states and completely topologically nontrivial bound states. Therefore, it is essential to distinguish whether a mid-gap state is topological or not via other properties.

The oscillation of zero-energy splitting is one of the unique features of partially coupled Majorana bound states. If a pair of Majorana bound states are separated by a distance $L$, which is comparable to the coherence length of the $p$-wave superconductor $\xi$, the coupling between the pair of Majorana bound states is finite and leads to a non-zero energy splitting $\Delta E$. The splitting can be estimated by: 
\begin{equation}
\label{EQ:split}
\Delta E \approx \frac{\hbar^2k_F}{m^*\xi}e^{-2L/\xi}\cos{(k_FL)},
\end{equation}
with $m^*$ the effective electron mass and $k_F$ the effective Fermi vector~\cite{Cheng2009, Sarma2012PRB}. Indicated by this equation, the Majorana splitting $\Delta E$ has an oscillating term $\cos{(k_FL)}$ and an exponential decay term $e^{-2l/\xi}$. The decay term is easy to understand because it reflects that the coupling is dominated by the effective length of the $p$-wave superconductivity $L/\xi$. The oscillatory term is also a direct consequence of the wave nature of Majorana. Here, we would like to relate it to the bottom-up picture shown in Fig.~\ref{Fig3} again. The available wave vectors for a one-dimensional object with length $L$ are given by $k_n=\frac{2\pi n}{L}$ ($n=0, \pm 1, \pm 2, ...$). Therefore, when $k_F$ is varied from $k_n$ to $k_{n+1}$, $k_FL$ is changed by $2\pi$ and results in an oscillating node. This is consistent with the picture where the merged sub-gap states in Fig.~\ref{Fig3} (b) will cross and split until they are pushed back by another interacting state at higher energy. Due to the parabolic dispersion relation, the oscillating period and amplitude will increase when $k_F$ is increased by tuning the chemical potential or the magnetic field [see Figs.~\ref{Fig5} (a-b)]. 

However, the oscillatory pattern in realistic experiments could be quite different. It is difficult to sweep one parameter and fix all of the other settings. For instance, when one sweeps gate voltage (chemical potential) to vary $k_F$, semiconductor-superconductor coupling $\Gamma$, superconductor coherence length $\xi$, effective $g$-factor, and spin-orbit coupling strength may all change~\cite{Antipov2018, Mikkelsen2018, Reeg2018, Dmytruk2018, Cao2019}, and lead to a more complex oscillating pattern. Reference~\cite{Sarma2012PRB} has also explained the delicate constant density-or-chemical potential problem and concluded that monotonically increase oscillations in most experimental setups would not be seen as Zeeman energy is varied.

In some of the semiconductor-superconductor experiments with emergent mid-gap states, ambiguous oscillating behavior has already been reported. For example, Ref.~\cite{Churchill2013} reported zero-bias features that oscillate in amplitude as a function of the magnetic field and gate voltage, though the soft-gap background made it difficult to track the sub-gap states. 

It is important to emphasize that, as the signature of partially coupled Majorana modes, the low-energy states have to be kept gapped while oscillating. Therefore, we have to characterize the oscillation in a cleaner sub-gap background. As shown in Fig.~\ref{Fig5} (c) from Ref.~\cite{Deng2016}, the gapped characteristic ``eye"-feature of the oscillating states as a function of gate voltage can be seen clearly in the hard gap. The characteristic ``eye"-shape oscillations have also been observed as a function of the magnetic field, as shown in Figs.~\ref{Fig5} (e).

In some cases, the characterization of oscillations in energy is limited by the measurement resolution (and the temperature/coupling broadening). Still, the oscillation can be reflected by the zero-bias conductance amplitude as a characteristic ``breathing" fine structure~\cite{Deng2018}, as shown by the gate and magnetic field dependent measurements in Figs.~\ref{Fig5} (d) and (f), respectively. The conductance amplitude breathing is simply because the overlapping of Majorana wavefunction will dramatically influence the Andreev-reflection amplitude at the normal probe-superconductor interface.

The oscillations discussed above are from devices with fixed physical lengths. In Refs.~\cite{Albrecht2016} and~\cite{Vaitiekenas2020}, the relations between oscillations and device lengths have been investigated. There, the lowest energy sub-gap states of hybrid islands are monitored by the Coulomb resonance peak space. The oscillations can be measured with a high resolution by averaging a large number of peak spaces. As shown in Figs.~\ref{Fig5} (h)-(i), the oscillation amplitude and device length follow an exponential decay trend.

It is worth emphasizing that the oscillation pattern or the length dependence should not be considered as a ``smoking-gun" signature of Majorana modes , but should only serve as evaluation criteria with other signatures. Trivial Andreev bound states or quasi-Majorana bound states~\cite{Prada2020, Moore2018, Vuik2019, Avila2019} could also introduce certain oscillation patterns.

\section{Quantization of Majorana conductance}

\begin{figure*}
\centering \includegraphics[width=18 cm]{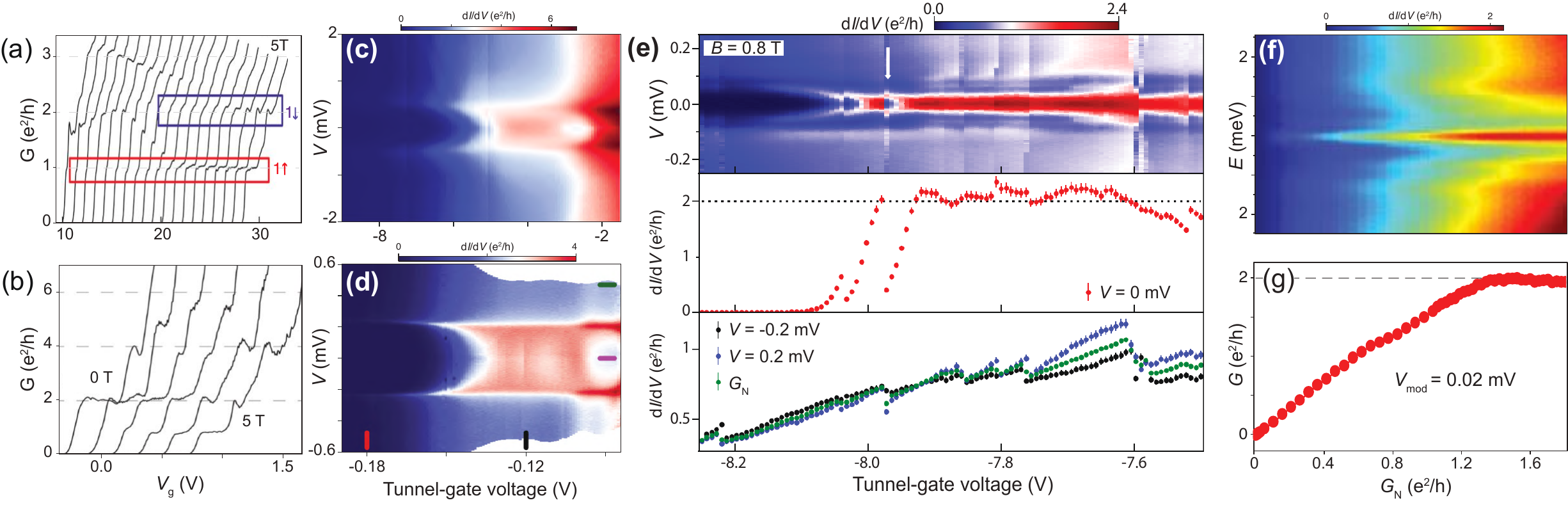}
\caption{\label{Fig6} Measurements of nearly quantized conductances. 
(a) and (b) Quantized conductance plateaus measured for InSb nanowire devices, with and without magnetic fields. Adopted from Refs.~\cite{Weperen2013} and \cite{Kammhuber2016}, respectively. 
(c) and (d ) Quantized conductance plateaus and Andreev conductance doubling measured for InSb-superconductor hybrid devices. Adopted from Refs.~\cite{Zhang2017} and \cite{ZhangH2021}, respectively. 
(e) Large conductance zero-energy peak of an InSb-Al nanowire device. Adopted from Ref.~\cite{ZhangH2021}.
(f) Nearly quantized conductance of a vortex state in FeTe\textsubscript{0.55}Se\textsubscript{0.45} superconductor. (g) Line-cut plot of the tunneling conductance of the zero-energy state. Adopted from Ref.~\cite{Zhu2020}.} 
\end{figure*}

Another spectacular signature of Majorana bound states is the quantization of the tunneling conductance measured from a Fermi lead~\cite{Law2009, Wimmer2011}. It was also considered to be a smoking-gun signature in tunneling spectroscopy measurement experiments. 

To understand the quantized Majorana conductance, we recall a normal mesoscopic conductor with a nanoscale constriction (the so-called quantum point contact), where the conductance will be quantized to $2ne^2/h$ (with $n=0, 1,2,3, ...$). This conductance quantization requires low temperature and ballistic transport. If the nano-constriction is replaced by two tunneling barriers that are closely separated, unity transmission occurs when the two tunneling barriers have equal transmission. In the perfect resonant regime, quantized conductance can also be achieved at resonance points. 

The perfect resonant tunneling is anticipated to always happen at the boundary of a normal lead and a Majorana zero-mode, giving rise to a quantized conductance. The symmetric particle-hole nature of Majorana zero-mode guarantees the equality between the transmission of the incoming particles (holes) to the lead and the transmission of the Andreev-conjugated holes (particles) to the same lead. That is to say, the normal lead close to the Majorana mode plays the role of both an electron lead and a hole lead simultaneously. Ballistic Andreev reflection will lead to a $4e^2/h$ tunneling conductance for each spin-degenerate mode (the BTK model~\cite{BTK1982}). For Majorana bound state, the spin degeneracy is lifted, and therefore the quantized Majorana conductance is $2e^2/h$. This conductance should remain precisely quantized even with disorders. However, quantized Majorana conductance requires the pair of Majorana modes sufficiently separated. This is because an overlap will cause the finite energy splitting given in Eq.~\ref{EQ:split}, which removes the particle-hole symmetry resulting from the zero-energy property of Majoranas. Transparency-independent quantized conductance plateau is thus a piece of substantial evidence for Majorana zero-mode.

In a practical experiment, the one-to-one correspondence between quantized conductance and Majorana bound state is unfortunately ambiguous. On the one hand, the tunnel barrier with a faulty profile, the temperature broadening effect, the multi-band effect, and other dissipations will all cause reductions to the quantized conductance in Majorana point contact structure~\cite{Pientka2012, Rainis2013, Liu2017-2}. On the other hand, a zero-energy state with a mundane origin is also possible to give a seemingly quantized conductance plateau along with a single tuning parameter~\cite{Moore2018, Pan2020, Yu2021}. Hence, the robustness of the quantized conductance has to be checked in high-dimensional parameter space.

Tunneling conductances measured in the first generations of Majorana devices are substantially lower than the expected $2e^2/h$ bar, even at low temperatures ($<50$~mK). Tremendous effort has been invested in reducing the dissipation in hybrid nanowire systems. For example, the hard-gap of full-epitaxy InAs/Al hybrid nanowires can significantly decrease disorder scattering in the Majorana channel. Nevertheless, it is difficult to make a clean tunneling barrier in one-dimensional InAs devices. This is because the conduction channels of InAs nanowires are naturally distributed at material surfaces and severely suffer from surface scatterings. 

To some degree, the scattering can be relieved in a lateral point contact geometry of a two-dimensional InAs/Al epitaxial material, in which quantized BTK conductance doubling can be obtained~\cite{Kjaergaard2016}. One dimensional hybrid wire supporting zero-energy state can be defined out of this two-dimensional system by selectively etching and gating~\cite{Suominen2017}. In Ref.~\cite{Nichele2017}, a zero-energy peak with an approximate $2e^2/h$ saturation conductance has been observed. 

In contrast to InAs nanowire, InSb nanowire has an intrinsic p-type doping surface, and hence its conducting channels are confined in the core region~\cite{Gul2015}. Electron transport is much less scattered in the InSb nanowire junction. As shown in Figs.~\ref{Fig6} (a) and (b), quantized conductance staircases from InSb tunneling devices are evident demonstrated~\cite{Weperen2013, Kammhuber2016}. With the high-quality ballistic constriction, Andreev conductance doubling has also been observed in metal-InSb-superconductor junctions [see Fig.~\ref{Fig6}(c) from Ref.~\cite{Zhang2017}]. With the development of full-epitaxial InSb/Al nanowire and shadow junction growing technology~\cite{Gazibegovic2017}, even better quality devices have been made. In these devices, the hard-gap spectrum in the low-transmission regime and the conductance doubling in the high-transmission regime have been both observed [Fig.~\ref{Fig6} (d)]. However, zero-energy states with convincing quantized conductance are still missing in hybrid nanowires so far, although large-conductance peaks can be observed in InSb-nanowire devices with the help of parameter-tuning~\cite{ZhangH2021}. 

Interestingly, nearly quantized conductance plateaus have been reported in a different but relevant system, i.e., vortex zero modes in iron-based superconductors~\cite{Chen2019cpl, Zhu2020}. Emergent zero-energy modes in the cores of ferromagnetic-superconductor vortices are probed by a scanning-tunneling-microscopy tip. The tunneling transmission can be varied by adjusting the tip-sample distance. As the tunneling conductance above the gap increases, the peak height of the zero-energy vortex mode is found to saturate at a value that is close to or equal to $2e^2/h$.

\section{Interacting Majorana with extra degrees of freedom}

So far, we have discussed the connotations of Majorana physics in one-dimensional hybrid devices by examining the emergence of zero-energy conductance peaks, the oscillations of the peaks, and their quantization. This section will extend the study of Majorana zero-modes to their interactions with some extra degrees of freedom.

The very first extra freedom is from the probing lead. For instance, in most of the experiments, the probes are Fermi leads with trivial density-of-state distributions. If the metallic probe can be replaced by a superconductor probe, thermal excitation can be greatly suppressed, and the measured conductance quantization should be more robust~\cite{Peng2015}. Another example is replacing the normal probe with a spin-polarized one. According to the theoretical prediction, for a Majorana end state with a specific local spin-polarization~\cite{Sticlet2012}, tunneling conductances from probes with different spin-polarizations could be quite different~\cite{He2014}. 

\begin{figure*}
\centering
\includegraphics[width=17 cm]{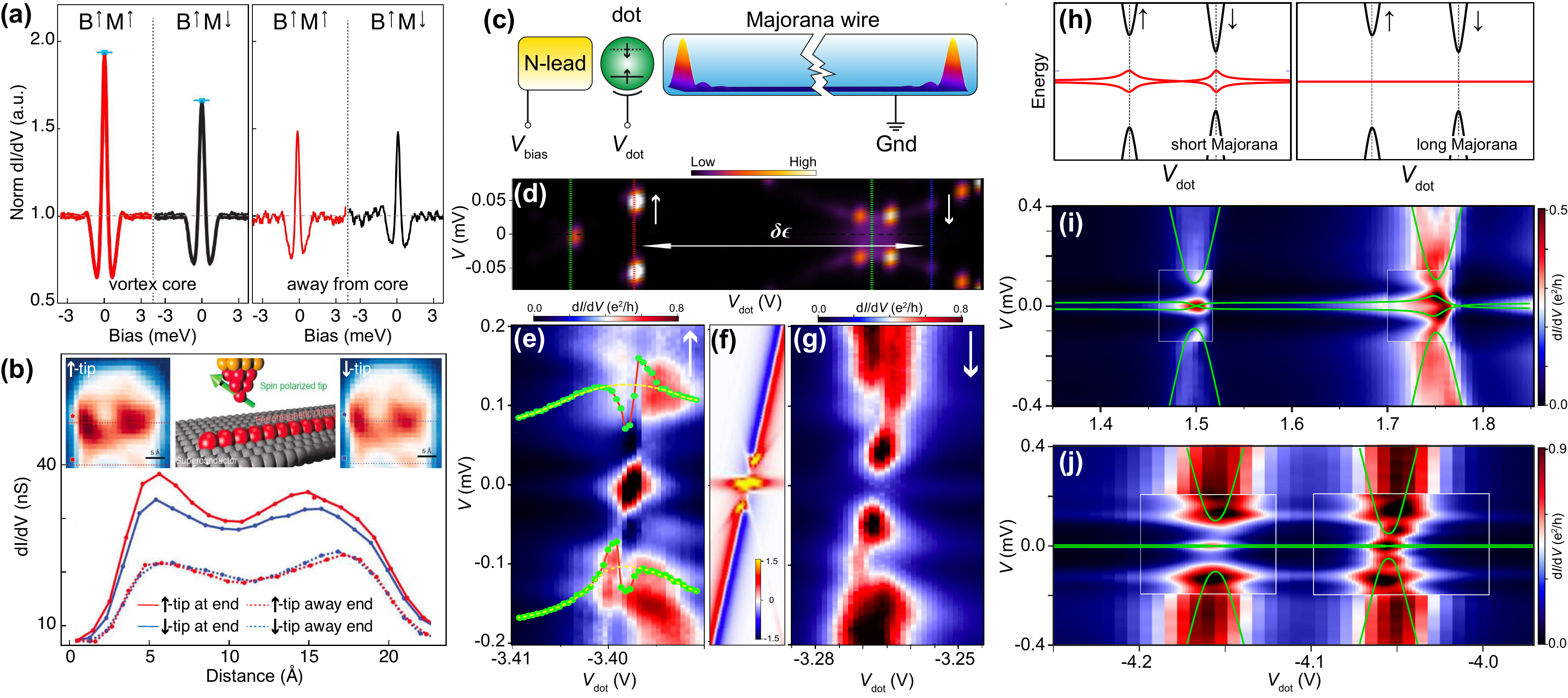}
\caption{\label{Fig7} Majorana bound states interact with extra degrees of freedom.
(a) Spin-selective Andreev reflection evidence observed in vortices of Bi\textsubscript{2}Te\textsubscript{3}/NbSe\textsubscript{2} heterostructure. Data reproduced Ref.~\cite{Sun2016}. 
(b) Spin-resolved zero-energy states as a signature of Majorana modes in a ferromagnet chain on a superconductor substrate. Data reproduced from Ref.~\cite{Jeon2017}. 
(c) Schematic of a coupled quantum dot-Majorana wire device. 
(d) Calculated tunneling conductance spectrum of a topological superconducting wire via a spin-polarized dot. The dot spin can be used as a tool to detect the spin polarization of the wire states. Data reproduced from Ref.~\cite{Chevallier2018}, with the dot spins reversed to match the negative Lande $g$-factor in the semiconductor. 
(e) Measured resonance tunneling between a spin-up dot level and the sub-gap states in an InAs/Al hybrid nanowire.
(f) Calculated conductance stability diagram of a dot-Majorana setup, where the fermion parity switching mechanism is considered. Data reproduced from Ref.~\cite{Leijnse2011}. 
(g) Similar to (e) but with spin-down level. Similar to (d), the dot spin in (e) and (f) demonstrate different selective features for the zero-energy state and finite-energy states.
(h) Calculated spectrum of coupled dot-Majorana wire setup. The dot can be used as a measure of the degree of Majorana nonlocality.
(i) and (j) Measured conductance stability diagram of coupled quantum dot-InAs/Al nanowire device. The extracted nonlocality degrees of the zero-energy are medium and high, respectively. (h)-(i) are adopted from Ref.~\cite{Deng2018}.} 
\end{figure*}

In Refs.~\cite{Sun2016}and~\cite{Jeon2017}, spin-polarized scanning-tunneling-microscopy tips are employed to probe iron atomic chains on superconductor and topological superconductor vortices, respectively. Both experiments show zero-energy states whose conductances strongly depend on the spin-polarization of the tip [Figs.~\ref{Fig7} (a) and (b)].

Similar experiments are also proposed in hybrid nanowire systems~\cite{Chevallier2018}. The spin injection can be realized using a spinful quantum dot between the normal lead and the hybrid nanowire as a spin filter. Such a quantum dot can form naturally or can be easily defined by electrical gates in the semiconductor constriction part. In the applied magnetic field (which causes phase transition in the hybrid nanowire), the spin-degeneracy of the quantum states in the quantum dot is lifted. We assume a spin-up level is aligned with the Fermi-level of the normal metal and the hybrid wire without losing generality. In this case, only spin-up electron transport is allowed without considering high-order cotunneling events [Fig.~\ref{Fig7} (c)]. Depending on the alignment details between the dot spin and the Majorana spin, the tunneling conductance to Majorana zero-modes can be greatly modified [Fig.~\ref{Fig7} (d)].

Measurements for normal metal-quantum dot-hybrid nanowire devices have been reported in Ref.~\cite{Deng2018}. As shown in Figs.~\ref{Fig7}(e) and (g), the conductance corresponding to the emergent zero-energy state highly depends on the spin of the quantum dot. At the same time, the conductance corresponding to the sub-gap states with higher energy shows a spin-dependence opposite to the zero-energy mode. This is in line with the prediction from Ref.~\cite{Chevallier2018}.

Besides providing spin as an interacting degree of freedom, the quantum dot can also serve as a single charge valve, which can be used to estimate the lifetime of Majorana parity~\cite{Leijnse2011}. The long parity-lifetime of Majorana will result in a sharp negative differential conductance along with one of the Coulomb diamond edges [Fig.~\ref{Fig7}(f)]. In Figs.~\ref{Fig7}(e) and (g), such a negative differential conductance pattern (after neglecting the background conductance) has been identified.

More importantly, a quantum dot coupled to the hybrid wire can be used to estimate Majorana nonlocality~\cite{Clarke2017, Prada2017, Schuray2017}. The coupled quantum dot can essentially drive a fine-tuned trivial zero-energy state off the sweep spot and split the zero-energy state. Indicated by Fig.~\ref{Fig7} (h), a low nonlocality Majorana bound state or a trivial Andreev bound state (red) will be strongly perturbed by a coupled dot level (black). In contrast, a highly nonlocal Majorana bound state will remain unchanged as it crosses the dot level. Such nonlocality measurements have been reported in Refs.~\cite{Deng2016, Deng2018}, see Figs.~\ref{Fig7} (i) and (j).

As proposed by many other theoretical works, the quantum dot is a versatile tool for Majorana verification~\cite{Liu2011, Zitko2011, Ruiz2015, Chirla2016, Schulenborg2020, Ricco2018} and application~\cite{Gharavi2016, Karzig2017, Malciu2018}, and this field is still actively researched currently. 

\section{Other experiments of interests}

There is plenty of other experimental progress of one-dimensional Majorana hybrid devices. We will briefly introduce some of them in this section but will not discuss the in-depth details.

\begin{figure*}[t]
\centering \includegraphics[width=18 cm]{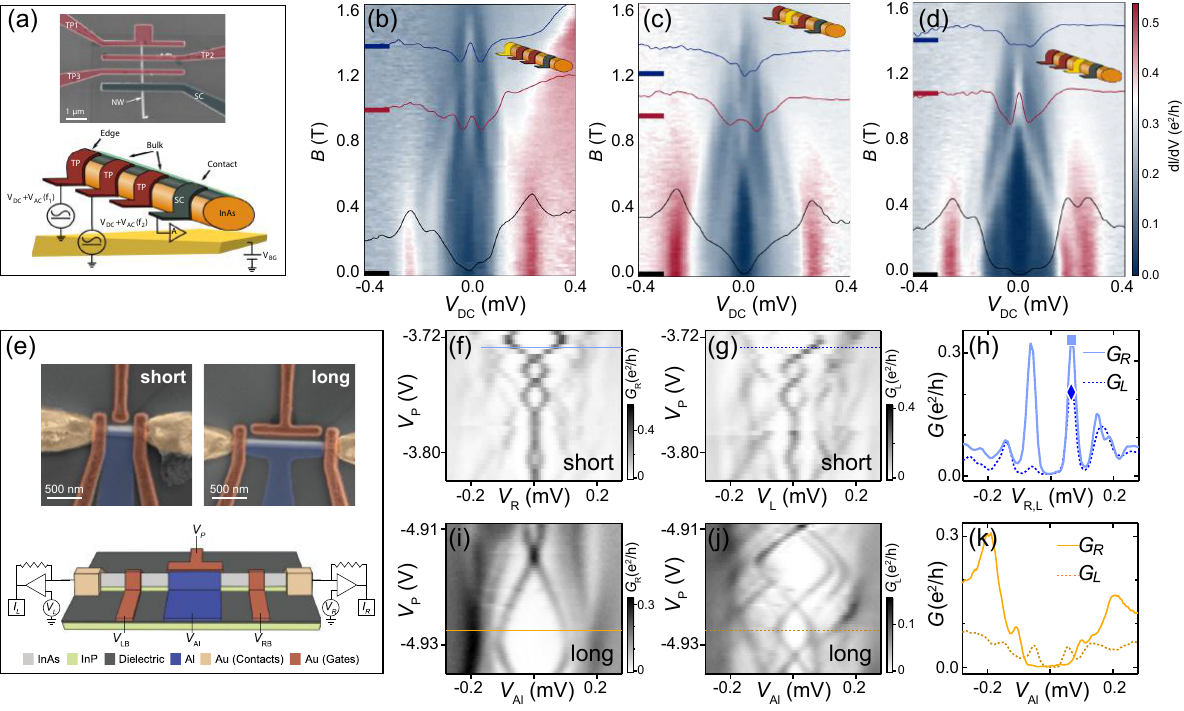}
\caption{\label{Fig8} Measurements of multi-terminal Majorana devices.
(a) Scanning-electron-microscope image and schematic of a four-terminal InAs/Al nanowire device. There are three normal metal probes and one superconducting grounding lead in the device.
(b)-(d) Measured tunneling conductance spectra of the device shown in (a), with different probes. The end lead probed an emerged zero-energy state in the magnetic field, while the bulk leads probed split states as a possible signature of the reopened superconducting gap. (a)-(d) Adopted from Ref.~\cite{Grivnin2019}. 
(e) Scanning-electron-microscope image and schematic of three-terminal selective-area-grown InAs/Al hybrid devices.
(f) Measured tunneling spectrum from the left side of the short device shown in (e).
(g) Same as (f), but measured from the right side of the short device. 
(h) Tunneling conductance comparison between the left-side and the right-side. The conductance spectra in (f)-(h) illustrate a clear end-to-end correlation.
(i)-(k) Same as (f)-(h), but measured for the long device shown in (e). No clear end-to-end correlation evidence is seen in the long device.
(e)-(k) are adopted from Ref.~\cite{Anselmetti2019}.} 
\end{figure*}

First of all, a few experiments use different mechanisms or strategies to relax the strict requirements for Majorana materials. The topological phase transition in the semiconductor-superconductor hybrid system is mostly driven by using a Zeeman field of order 1T. Usually, a strong applied magnetic field is incompatible with the superconductor device. In Refs.~\cite{Fornieri2019, Ren2019, Yang2019}, phase-controlled topological superconductivity signatures in two-dimensional material systems have been reported. The critical fields for observing zero-energy states are significantly reduced when the superconductor junctions are biased at $\phi\approx\pi$. Similar but different, it is also demonstrated that quantized vortices with odd-winding numbers in InAs/Al core-shell nanowires can host flux-induced Majorana mode~\cite{Vaitiekenas2020}. Again, it should be noted that the flux can also give rise to trivial zero-energy states mimicking Majorana modes in these core-shell wires~\cite{Valentini2020}. Phase or flux controlled topological superconductivity still requires applying an external magnetic field (though the critical field is reduced). In Ref.~\cite{Vaitiekenas2020-2}, zero-field topological superconductivity is realized via synthesizing semiconductor, superconductor, and the ferromagnetic insulator together. In this setup, the time-reversal symmetry of the induced superconductivity is broken by the magnetic exchange effect from a EuS layer in the close vicinity. Similarly, it is found that ferromagnetic microstructures with proper magnetic textures can induce a strong spin-orbit field and a Zeeman field for materials with weak intrinsic spin-orbit interactions~\cite{Desjardins2019}.

Secondly, devices with multi-terminals have been investigated recently. Most of the devices mentioned above are two-terminal devices, i.e., one probe terminal and one grounding terminal. In this geometry, the probe lead can only peek at Majorana zero-mode at one end. Information about the Majorana at the opposite and the bulk p-wave superconductor is, however, still missing. In Ref.~\cite{Grivnin2019}, the tunneling conductance at the wire-end and the conductance probed in the middle of the wire are monitored simultaneously. A zero-energy state at the end and a reopened gap signature in the middle gap are observed [Figs.~\ref{Fig8} (a)-(d)]. End-to-end correlation measurements are also performed in selective-area-grown material devices~\cite{Anselmetti2019, Puglia2020}. In Ref.~\cite{Anselmetti2019}, the different cross-correlations are found for devices with various lengths [Figs.~\ref{Fig8}(e)-(k)], while the nonlocal measurements in Ref.~\cite{Puglia2020} reveal the lack of clean topological superconductivity in the relevant system. A clear end-to-end correlation of Majorana conductance has not been reported up to now. 

Direct nonlocality measurement through end-to-end conductance correlation is difficult. This is because the Majorana mode extending over the entire nanowire is still quite rare. It needs long defect-free material and uniform parameter tuning. Also, an end-to-end correlation measurement normally requires a middle probe as a common drain end. For one-dimensional epitaxy InAs-Al or InSb-Al nanowires, the middle drain probe has to be made in the aggressive etching-evaporation way, which will easily induce defects in the hybrid segment. 

Other interesting experiments, like phase-diagram mapping~\cite{ChenJ2017, ShenJ2020} and Aharonov-Bohm interferometer~\cite{WhiticarAM2020}, for example, will not be addressed here. 

\section{Outlook and challenges }

As a growing field nowadays, experimental research on Majorana hybrid nanowire devices has more opportunities and challenges. As we mentioned previously, conclusive evidence of non-Abelian Majorana zero-mode existence is yet to be found, despite the observations of zero-energy state features reviewed above. Thus, it remains quite challenging to realize the first Majorana qubit.

\begin{figure*}
\centering \includegraphics[width=16 cm]{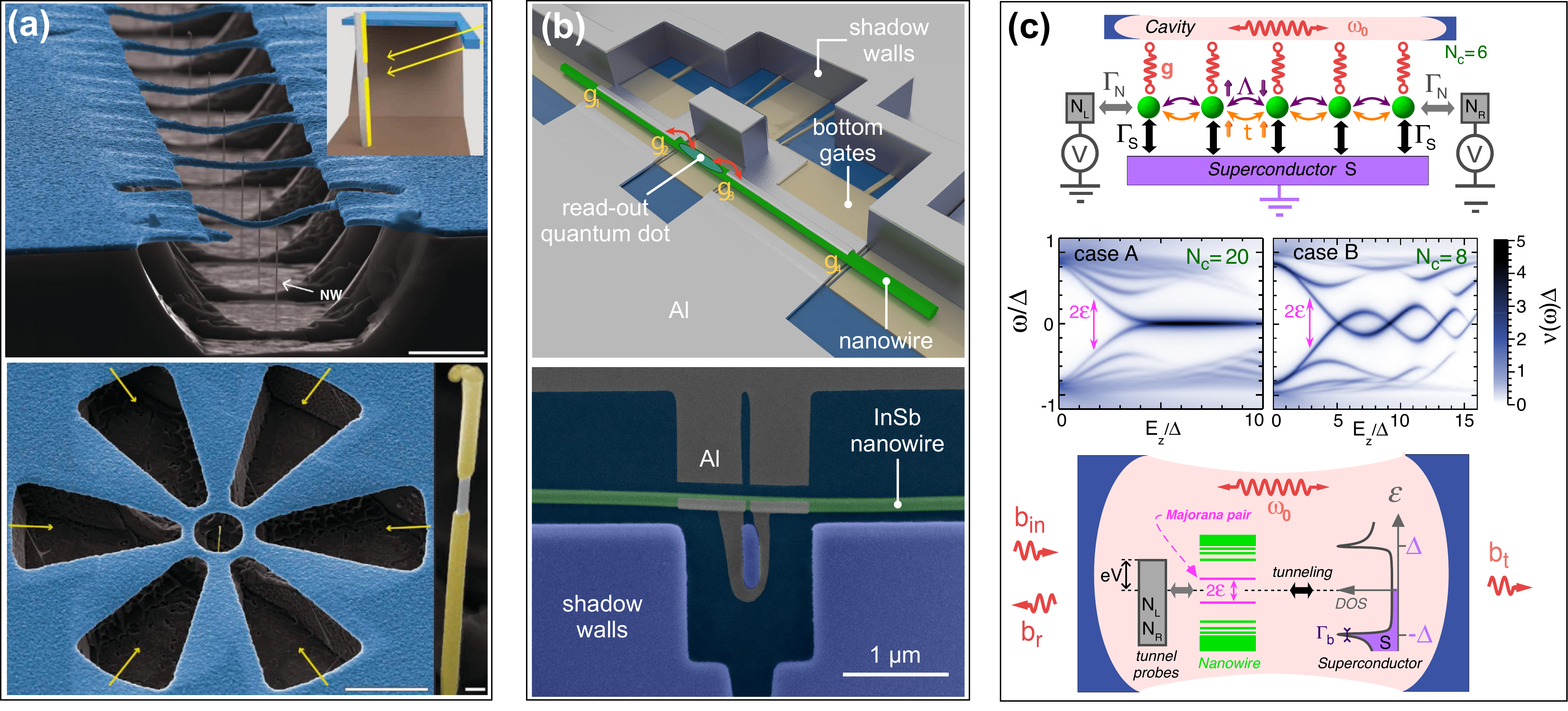}
\caption{\label{Fig9} Advanced material growth, fabrication and measurement. (a) Shadow epitaxy of \emph{in situ} hybrid material grown. Adopted from Ref.~\cite{Carrad2019}. (b) Shadow-wall lithography and single-shot device fabrication of hybrid devices. Adopted from Ref.~\cite{Heedt2020}. (c) Microwave cavity detection of Majorana zero mode. Adopted from Ref.~\cite{Dartiailh2017}.} 
\end{figure*}

Here, we would like to briefly discuss the main hurdles for the 0 to 1 breakthrough of hybrid nanowire topological qubit and the possible solutions in the future.

\bt{Materials}. Material probably is the most fundamental and the most critical element for Majorana zero mode~\cite{Wen2021}. However, the current materials are still not good enough. For example, the electron mobility of the semiconductor nanowires used in the Majorana experiments is low, which is in the order of a few thousand cm\textsuperscript{2}/Vs. As a reference point, the electron mobility in high-quality GaAs/AlGaAs two-dimensional electron gas systems can reach 10 million\textsuperscript{2}/Vs~\cite{Manfra2014}. It is of course not fair to compare a one-dimensional nanowire to two-dimensional films due to the intrinsic surface scattering issue of the one-dimensional system. However, in Ref.~\cite{Gul2015}, it is shown that the electron mobility is super-sensitive to the surface condition and that solely increasing the evacuation time of the nanowire device can significantly improve the electron mobility. Long evacuation can help remove the adsorption of molecules to the nanowire surface and the substrate, but it is not easy to reduce more intrinsic adsorptions, like oxidations or other dopants on the surface. Also, growth defects, e.g., stacking faults, lattice mismatch, unintentional doping, etc., will limit the formation of high-quality topological states. 

For the hybrid system, good semiconductors and good superconductors are equally important. For semiconductors, the material growers are still improving the growth of larger spin-orbit interaction and larger $g$-factor materials. For instance, band engineering of InSb-InAs compound nanowires, i.e., InAs\textsubscript{\emph{x}}Sb\textsubscript{\emph{1-x}}, could further increase the spin-orbit interaction~\cite{Sestoft2018}. Also, materials beyond one-dimensional systems, such as two-dimensional electron gas materials with epitaxial superconductor films~\cite{Shabani2016, Nichele2017, Fornieri2019} and network structures~\cite{Gazibegovic2017, Krizek2018, SAGprl2018, Lee2019, Aseev2019} have been researched. They can avoid wire transferring induced damages and form more complex structures. For superconductors, large gap epitaxial superconductors, like Sn, Pb, Ta, Nb, and V, etc., are under intensive investigation~\cite{Khan2020, Pendharkar2021}. Moreover, the semiconductor-superconductor interface engineering is critical for Majorana nanowires and requires delicate growth parameter tuning. Meanwhile, the development of ferromagnetic material/semiconductor/superconductor could benefit magnetic-field-free Majorana generating and operation~\cite{Vaitiekenas2020-2}. At last, we should always keep open to the development of exotic materials that have promising topological properties~\cite{Zhang2019n,Vergniory2019n,Tang2019n}. 

\bt{Devices}. Secondly, there is plenty of room for improvement in device fabrications. The traditional fabrication process of Majorana devices normally includes material transferring, dry/wet etching, e-beam/UV lithography, and metal deposition/lift-off. Contaminations and defects can be induced into the materials during these fabrication processes and prevent the formation of high-quality topological superconductivity. The device also suffers from a huge parameter tuning inhomogeneity problem. Due to the electrical field and/or the magnetic field screening effect, chemical potential and Zeeman energy tuning are not uniform. 

Therefore, novel fabrication technology that can avoid damaging material needs to be developed. One exciting direction is the \emph{in situ} device fabrication or single-shot fabrication~\cite{Carrad2019, Heedt2020, Borsoi2020}, which can move part of the device patterning process into the vacuum chamber and reduce defects. The development of resistless stencil lithography~\cite{Vazquez-Mena2015, Schuffelgen2019} could help achieve high-quality topological devices.  It is also worth looking outside the box and taking advantage of technologies from other fields, like employing spintronic arrays to control local magnetic fields~\cite{Huang2020}.

\bt{Measurements}. Thirdly, measurement methods beyond electron-transport are worth exploring. In most of the hybrid material devices, the tunneling electrodes are invasive contact. This kind of probe will induce device quality reduction during fabrication and will be quasiparticle poison sources~\cite{Budich2012, MenardPRB2019}.

New measurement methods are under investigation these years, including the microwave spectroscopy~\cite{Woerkom2017, Tosi2019}, the scanning SQUID microscopy~\cite{Spanton2017, Hart2019}, the dispersive readout method~\cite{Kringhoj2020}, and the fast charge sensing~\cite{Razmadze2019, Zanten2020}. Some of these are non-invasive measurements, which are important for qubit-wise analysis. 

More perspective and outlook works can see, for example, Refs.~\cite{Zhang2019, Frolov2020} on transport device design and material synthesizing. 

We believe that the profound impact on condensed matter physics and the enormous potential application values of Majorana zero-mode will attract more and more investment, and more fruitful Majorana experiment works can be expected.

\textbf{Acknowledgment.} This work was supported by the National Natural Science Foundation of China (NSFC) (11904399) and the Open Research Fund from State Key Laboratory of High Performance Computing of China (HPCL) (Grant No. 201901-09). M.T.D acknowledges the support from China Great-wall Quantum Laboratory.


\end{document}